\documentclass[aip, rsi, amsmath, amssymb, reprint]{revtex4-1}
\usepackage{graphicx}
\usepackage{dcolumn}
\usepackage{bm}

\begin{document}

\preprint{AIP/123-QED}
\title{Analyses of Scissors Cutting Paper at Superluminal Speeds}
\author{Neerav Kaushal}
\email{kaushal@mtu.edu}
\author{Robert J. Nemiroff}%
\email{nemiroff@mtu.edu}
\affiliation{ Department of Physics \\ Michigan Technological University \\ 1400 Townsend Drive, Houghton, MI 49931}
\date{\today}

\begin{abstract}
A popular physics legend holds that scissors can cut paper with a speed faster than light. Here this counter-intuitive myth is investigated theoretically using four simple examples of scissors. For simplicity, all cases will involve a static lower scissors blade that remains horizontal just under the paper. In the first case, the upper blade will be considered perfectly rigid as it rotates around and through the paper, while in the second case, a rigid upper blade will drop down to cut the paper like a guillotine. In the third case, the paper is cut with a laser rotating with a constant angular speed that is pointed initially perpendicular to the paper at the closest point, while in the fourth case, the uniformly rotating laser is pointed initially parallel to the paper. Although details can be surprising and occasionally complex, all cases allow sections of the paper to be cut faster than light without violating special relativity. Therefore, the popular legend is confirmed, in theory, to be true.
\end{abstract}

\maketitle

\section{\label{sec:level1}Introduction}

A common expression is that nothing can travel faster than light -- but this is not strictly true. Shadows, for example, are a well known counter example \cite{1995iqm..book.....G,Nemiroff2}. Conversely, it is common lore \cite{web:lang:stats7,web:lang:stats5,web:lang:stats1,web:lang:stats6} in physics that the vertex of a pair of scissors can exceed $c$ but no comprehensive pedagogical account of this exists. In this work we will explore the truth of this physics legend by analyzing four specific examples.

Special relativity states that any object possessing a finite mass cannot be accelerated from below the speed of light to above, because then its momentum and energy would diverge \cite{1905AnP...323..639E}. It is also well documented that locally created information cannot be transmitted faster than light. Light, neutrinos, relativistic particles, and gravitational waves; are all limited to travel in vacuum at, or very near, $c$. 

In this paper, it is shown that the cutting vertex of scissors can, in theory, move faster than light, relative to the inertial frame of the paper. In all of the example cases considered here, the upper edge of the lower scissors blade will be considered to lie just under the paper and directly along the $x$-axis, where $y$ is always zero. This axis extends along the paper and also along the line on which the paper will be cut. The thickness of both scissors blades will not be considered important. Therefore, the cases considered are actually variations of the geometry and movement of the upper blade.

Sections II and III will define and theoretically analyze four example cases. In the first two cases, the upper blade will be considered a rigid material. In Section II.A the upper blade will be considered to rotate about a pivot, while in Section II.B it will be considered to drop like a guillotine. In the second two cases, the upper scissors blade will be considered to be a uniformly rotating laser. In the third case analyzed in Section III.A, this laser will point initially at a direction perpendicular to the paper it cuts, while in the fourth case, in Section III.B, the laser will point initially parallel to the paper. Section IV summarizes the findings and gives some discussion.

\section{\label{sec:level1}Rigid Blade Scissors}

\subsection{\label{sec:level2}Rotating Upper Blade}

The case of a rigidly rotating upper scissors blade will be considered first. The upper blade starts by spanning the $y$-axis while being able to pivot at the position of its hinge, below the $x$-axis, at $(x,y) = (0,-L)$. Here $L$ is called the ``hinge distance". The blade rotates clockwise about the hinge starting at $t=0$. The upper scissors blade will be considered to close with a constant angular velocity $\omega$ on the fixed lower blade. Therefore, the point of intersection of the two blades, the vertex, being confined to the $x$-axis, moves from the origin towards increasingly positive $x$ at speed $v$. 
    
        \begin{figure}[!htb]
            \centering
            \hspace*{-2.0cm}
            \includegraphics[scale=0.32]{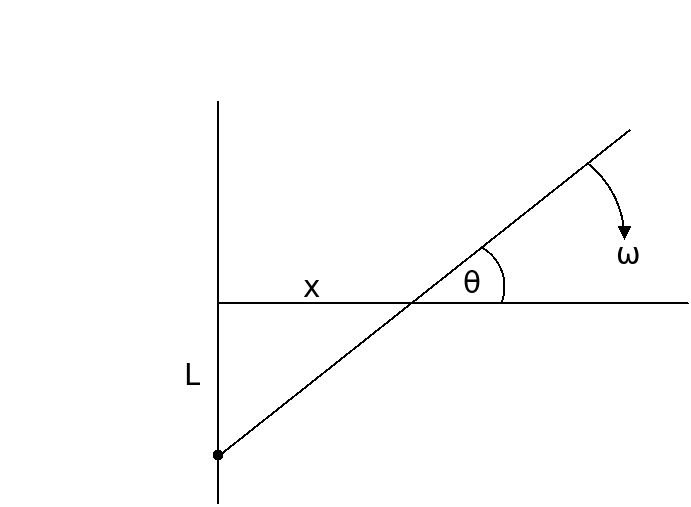}
        \caption{(Case II A) A scissors is depicted with hinge distance $L$ and rigid upper blade rotating about the hinge with constant angular velocity $\omega$. The angle the upper blade makes with the $x$-axis is $\theta$, while the distance to the vertex -- the place where the paper is being cut -- is designated $x$.}
            \label{fig:1}
        \end{figure}
    
    The upper edge of the lower blade and the lower edge of the upper blade are the only edges considered here to be important. Each edge is described by a line, and the vertex is the point where these lines intersect. The scissors is assumed to be entirely rigid so that when the upper blade closes with constant angular velocity $\omega$ on the lower blade, all of it translates at once, without flexing. The geometry is depicted in Figure~\ref{fig:1}. 
    
\begin{figure}[ht]
            \centering
            \hspace*{-0.5cm}
            \includegraphics[scale=0.55]{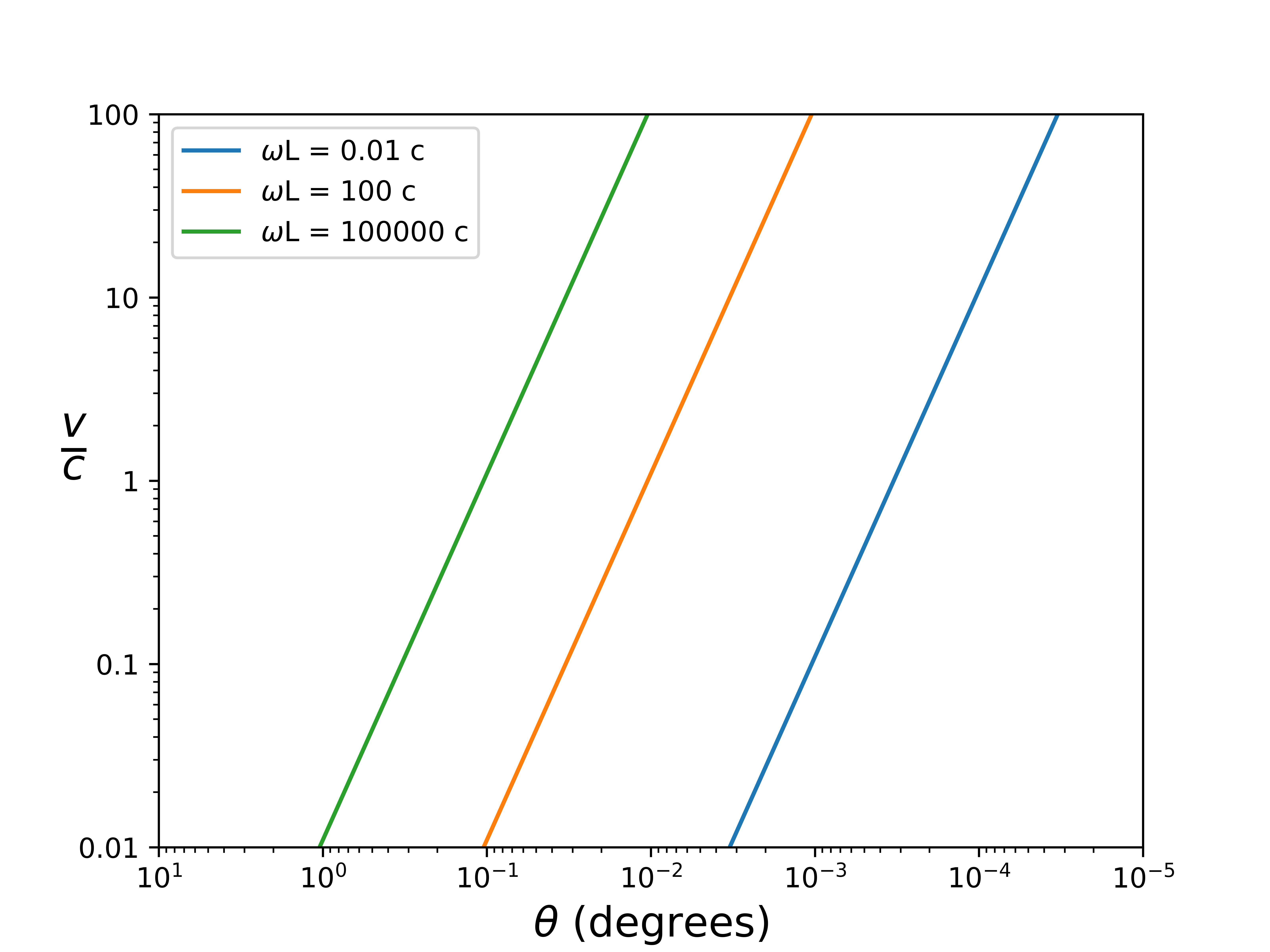}
            \caption{(Case II A) The velocity at which the vertex advances increases with decreasing angle between the blades. The angle at which this velocity exceeds the speed of light $c$ depends on the hinge distance of the scissors and the speed of rotation $\omega$ of the upper blade.}
            \label{fig:2}
        \end{figure}

    Simple Euclidean geometry shows that 
    \begin{equation}
    x = L \ \cot \theta .
    \end{equation}
    The velocity $v$ with which the vertex advances in the positive $x$ direction is given by
    \begin{equation} \label{EqV}
        v = \omega L \ \sin^{-2} \theta .
    \end{equation}
    The vertex velocity varies with the angle between the blades as shown in Figure~\ref{fig:2} .
   
    The vertex moves towards increasingly higher values of $x$ with increasing speed. After a finite time, the scissors will close completely, meaning that the upper scissors blade will become parallel to the lower blade. This will occur at time $t_{close}= (\pi/2)/\omega$. Considering both scissors blades to be infinitely long, the vertex will cross the entire {\it infinite} length of the positive $x$-axis in this finite time. Clearly, to accomplish this, it must approach an infinite speed. Therefore, the vertex speed will eventually exceed the speed of light. 
    
    The value of $\theta$ where the vertex speed $v$ exceeds $c$ will be called $\theta_c$ and can be found by setting Eq. (\ref{EqV}) equal to $c$. Straightforward algebra shows that
    \begin{equation} \label{EqTheta}
        \theta_c = \arcsin \sqrt{\frac{\omega L}{c}} .
    \end{equation}
    Similarly, the value of $x$ where the vertex speed $v$ exceeds $c$ will be called $x_c$ and can be found by integrating Eq. (\ref{EqV}). The result is 
    \begin{equation} \label{xc}
        x_c = \sqrt{\frac{L}{\omega} (c - \omega L)} .
    \end{equation}
    Last, the time $t_c$ when the vertex exceeds $c$ can be simply found from Eq. (\ref{EqTheta}) as
    \begin{equation}
        t_c = { ({\frac{\pi}{2} - \theta_c}) \over {\omega} } .
    \end{equation}
    Clearly, higher angular speeds $\omega$ create proportionally lower $c$-crossing times. This type of scissors can cut a piece of paper superluminally even with finite values of angular velocity. 
    
    The strict rigidity of the rotating scissors, however, makes this case unphysical in theory but there are possible ways to get around this which are discussed in detail in the last section.

\subsection{\label{sec:level2}Falling Guillotine}

\begin{figure}[ht]
            \centering
            \hspace*{-0.6cm}
            \includegraphics[scale=0.32]{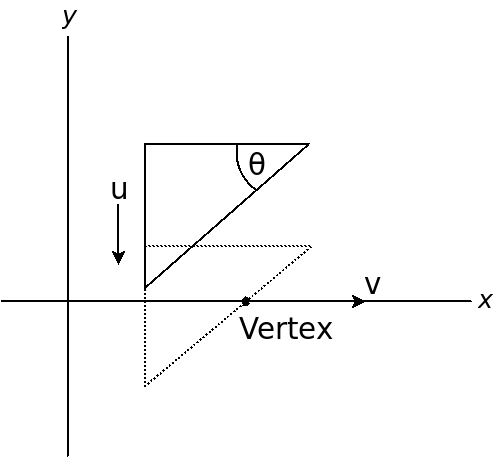}
            \caption{(Case II B) A guillotine-type scissors blade moves downwards so that its vertex moves along the $x$ axis.}
            \label{fig:4}
        \end{figure}

Another form of scissors is a guillotine -- a falling rigid upper scissors blade that makes a constant angle $\theta$ with the $x$-axis. The guillotine does not rotate -- it moves down at a constant ``blade speed" $u$ in the negative $y$ direction. For didactic purposes, the guillotine is considered to be infinitely long, and only the lower edge of the guillotine blade is important. This edge will be considered perfectly straight \cite{web:lang:stats2}. As before, the paper to be cut is laid across the $x$-axis. The vertex in this case moves along the positive $x$-axis with velocity $v$ as shown in Figure~\ref{fig:4}.
    
    The velocity $v$ of vertex of the inclined surface of guillotine and the paper is given by 
    \begin{equation}
        v = u \ \cot \theta .
    \end{equation}
    For a given downward speed $u$, the critical guillotine angle below which the blade cuts the paper faster than light is $\theta_c = \cot^{-1} (c/u)$. Slower guillotine speeds need a shallower angle to cut the paper faster than light.
    
    If the guillotine speed $u$ is constant, then the speed of the vertex $v$ is also constant. Therefore, this guillotine scissors either always cuts the paper superluminally, or it never does. 
    
\begin{figure}[ht]
            \centering
            \hspace*{-0.5cm}
            \includegraphics[scale=0.55]{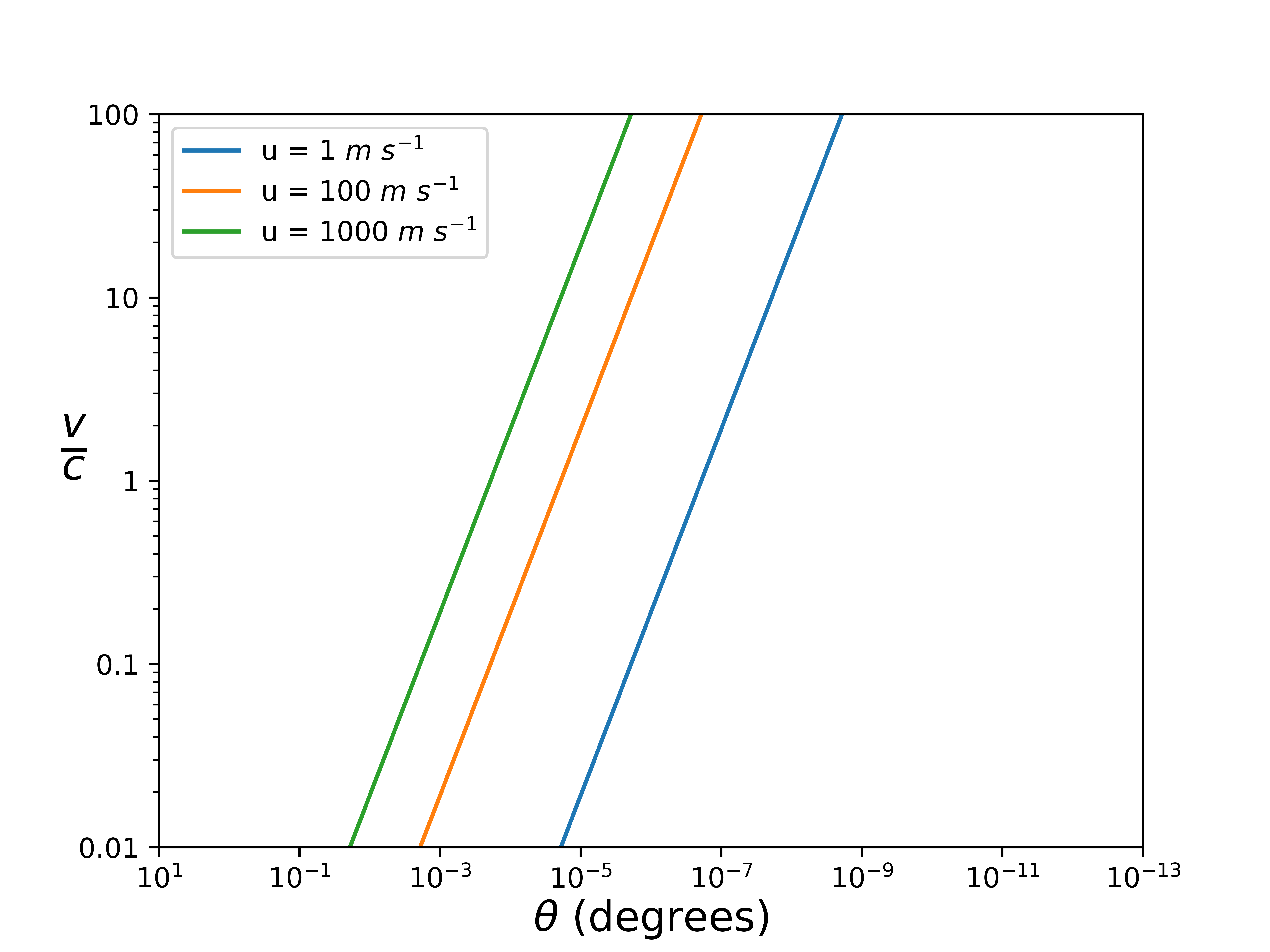}
            \caption{(Case II B) A plot of vertex velocity versus the guillotine angle for different blade speeds.}
            \label{fig:5}
        \end{figure}

    The variation of the vertex velocity with the guillotine angle for three different blade velocities is shown in Figure~\ref{fig:5}. Note that the speed of the vertex varies linearly with the guillotine angle. For example, for a guillotine dropping with a speed of $u = 1$ m s$^{-1}$, the vertex reaches $c$ at an angle of $1.91 \times 10^{-7}$ degrees with respect to the horizontal. It is not feasible in practice to create a guillotine with such a small angle because atomic scales -- too fine to control -- become involved. 

\section{\label{sec:level1}ROTATING LASER SCISSORS}

\subsection{\label{sec:level2}Starting Orientation: Beam perpendicular to the paper}

A different type of scissors is now considered: a laser scissors. Cutting paper with a laser is now relatively common in industry as lasers are inexpensive and can be pointed with high accuracy \cite{web:LaserPaperCutter}. Furthermore, in terms of the physics, laser beams do not have rigidity concerns that regular material scissors have shown in the previous analyses, yet the beam has well defined characteristics that can be analyzed. For didactic purposes, this laser beam will be considered to be perfectly collimated -- when stationary, the laser emits a thin stream of photons linearly out from the its base. The scissors vertex will be considered as the intersection point of the laser beam with the paper. It will be assumed that whenever laser light touches the paper, it cuts the paper, even though this may be impractical when far from the laser source. As before, the paper will be considered laying flat in the $x-z$ plane with the laser constrained to cut the paper along the positive $x$ axis. 
    
\begin{figure}[ht]
            \centering
            \includegraphics[scale=0.32]{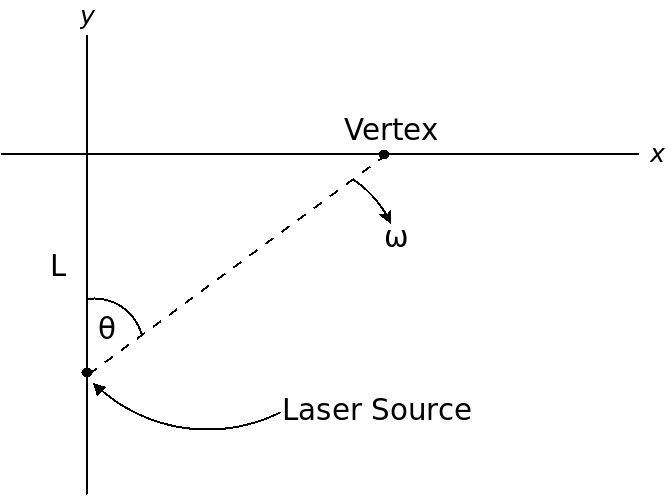}
            \caption{(Case III A) A laser source at $(0,-L)$. At $t=0$, the laser is switched on, and given an angular velocity 
            $\omega$ so that the vertex moves along $+x$-axis.}
            \label{fig:7}
        \end{figure}

    Consider now a specific laser emanating light from $(x,y) = (0,-L)$ towards the positive $y$-axis. For simplicity, the laser is considered hinged at the point of light emanation, and rotated about this hinge clockwise at constant angular velocity $\omega$. Therefore, the laser cuts the paper such that the vertex moves away from the origin toward positive $x$ values, as shown in Figure~\ref{fig:7}. 
        
    At $t=0$, the laser is switched on so that the first laser photon travels towards the origin. Also at $t=0$, the laser begins pivoting about the hinge at angular velocity $\omega$. There will be a single spot of light cutting the paper along the $x$-axis while the laser is being rotated from 0$^{\circ}$ through 90$^{\circ}$ with respect to its initial orientation. The time it takes for the laser to rotate through an angle $\theta$ is given by
    \begin{equation}
        t_{rotate} = \frac{\theta}{\omega} .
    \end{equation}
    Once the laser has turned to angle $\theta$, the photons emanating from it will take a finite time to traverse the space between the laser and the paper, ultimately reaching the paper at angle $\theta$ on the $x$-axis. This time is given by 
    \begin{equation}
        t_{traverse} = \frac{L}{c \  \cos\theta} .
    \end{equation}
    The total time from $t=0$ to when a laser photon touches and cuts the paper at angle $\theta$ is given by the sum of these two times such that
    \begin{equation}
        t_{total} = t_{rotate} + t_{traverse} = \frac{\theta}{\omega} + \frac{L}{c \ \cos\theta } .
    \end{equation}
    When the laser finally points at 90$^{\circ}$, the beam takes infinite time to reach infinitely far down paper.
    
\begin{figure}[ht]
            \centering
            \hspace*{-0.5cm}
            \includegraphics[scale=0.57]{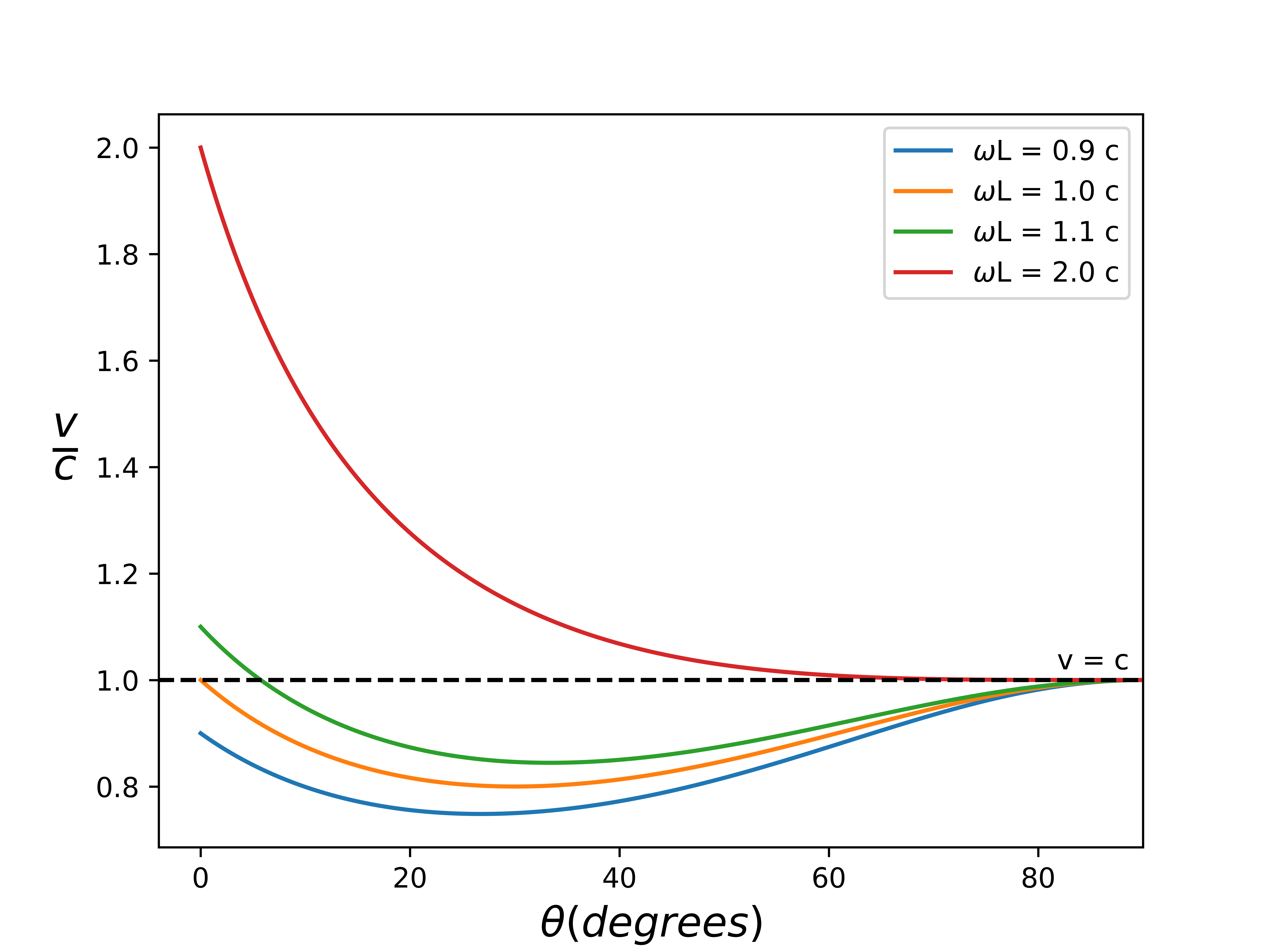}
            \caption{(Case III A) Velocity of vertex $v$ vs the angle $\theta$ with respect to the vertical}
            \label{fig:8}
        \end{figure}

    Following Nemiroff \cite{Nemiroff1}, the speed of this real beam spot or vertex along the surface of paper, $v$ is given by
    \begin{equation}
        v = \frac{L}{\cos\theta} \frac{d \theta} {dt_{total}} 
          = \frac{\omega\ c\ L}{c\ \cos^{2}\ \theta + \omega\ L\ \sin \theta} .
    \end{equation}
    A plot of $v$ versus $\theta$ is shown in Figure~\ref{fig:8}.
    At the origin, the initial vertex speed is $v (initial) = \omega \ L$. Note that this speed is {\it not} constrained to be slower than light: if either $\omega$ or $L$ is fixed, the other variable can be increased so that $v (initial) > c$. 
    
    The behavior of $v$ is a surprisingly complicated function, dropping initially as $t_{traverse}$ increases relatively rapidly for small $\theta$. The fastest $v$ will occur at the origin unless $v (initial) < c$. If $v (initial)$ is sufficiently large, then $v$ will drop below $c$ before recovering at large $x$. Regardless of $v (initial)$ and its low $\theta$ behavior, the scissors vertex speed $v$ will always asymptote to $c$ as $x$ goes to infinity, far down the paper because there the laser photons travel nearly parallel to the paper.

\subsection{\label{sec:level2}Starting Orientation: Beam parallel to the paper}

\begin{figure}[ht]
        \centering
        \hspace*{-0.5cm}
        \includegraphics[scale=0.40]{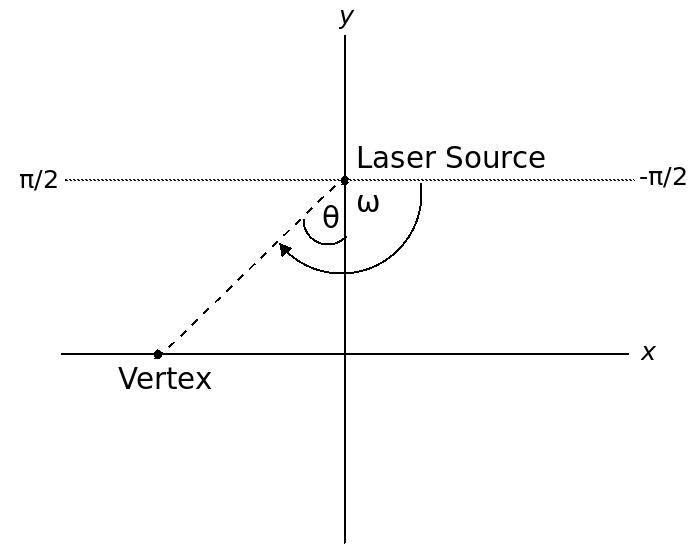}
         \caption{(Case III B) A laser source residing at $(0,+L)$, at $t=0$ is given an angular velocity \(\omega\).}
        \label{fig:10}
    \end{figure}

Now consider a laser emanating light from $(x,y) = (0,L)$ in the positive $x$ direction, parallel to the $x$ axis. As before the laser is hinged at the point of light emanation and the paper is laid across the $x$ axis in the $x-z$ plane. Now, however, the laser will {\it not} start cutting the paper at the origin, but at a point down the positive $x$ axis. The initial geometry is shown in Figure~\ref{fig:10}. 
    
    If the previously analyzed cases of scissors cutting paper were more surprising and complex than one might naively expect, then this case is even more surprising, and perhaps even bizarre! Here, the total time for the beam to illuminate and hence cut through a point at angle $\theta$ is given by the sum of the time it takes for the laser to rotate to angle $\theta$ and the time it takes for light to go from the laser to the point on $x$-axis at angle $\theta$, such that \cite{Nemiroff1}
    \begin{equation}
        t_{total} = t_{rotate} + t_{traverse} = \frac{\theta + \pi/2} {\omega} 
                   + \frac{L}{ c\ \cos\theta} .
    \end{equation}
    
\begin{figure}[ht]
            \centering
            \hspace*{-0.5cm}
            \includegraphics[scale=0.48]{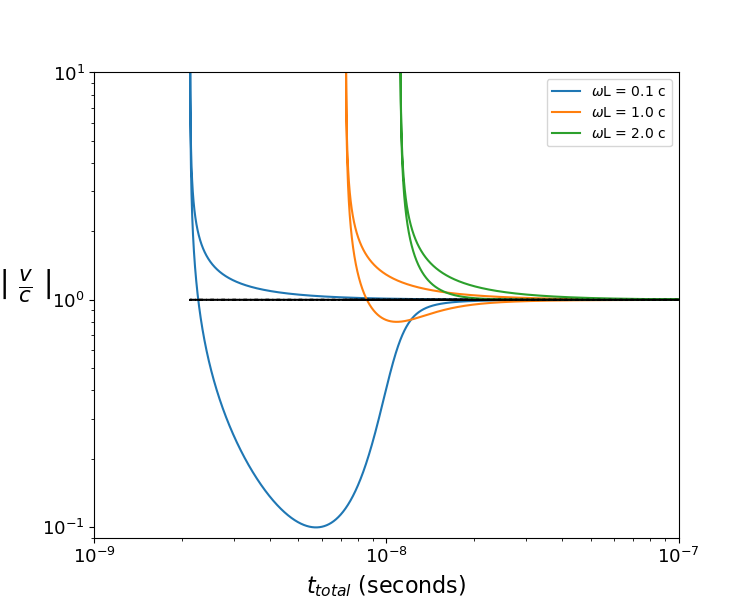}
            \caption{(Case III B) Variation of $\mid \frac{v}{c}\mid$ vs $t_{total}$ of the laser scissors vertex.}
            \label{fig:12}
    \end{figure}

    The beam starts at $t = 0$ by pointing along $\theta = -\pi/2$, parallel to the positive $x$-axis, and then rotates at constant angular speed $\omega$ so that $\theta$ increases through zero and ends when $\theta = +\pi/2$, where it points along the negative $x$-axis. When the laser is pointing towards the negative $\theta$, then $\sin{\theta}$ will be negative. The speed of the vertex along the surface of paper or the speed at which paper is cut, $v$ is given by 
    \begin{equation}
        v = \frac{L}{\cos\theta} \frac{d\theta}{dt_{total}} 
        = \frac{\omega\ c\ L}{c\ \cos^{2} \theta + \omega\ L \sin \theta} .
    \end{equation}
    When $v$ is positive, the location where the laser beam hits the $x$-axis -- the vertex -- moves in the direction of positive $x$, while a negative $v$ indicates vertex motion in the negative $x$ direction. A plot of $\mid \frac{v}{c}\mid$ vs $t_{total}$ is shown in Figure~\ref{fig:12}
    
\begin{figure}[ht]
            \centering
            \hspace*{-0.4cm}
            \includegraphics[scale=0.51 ]{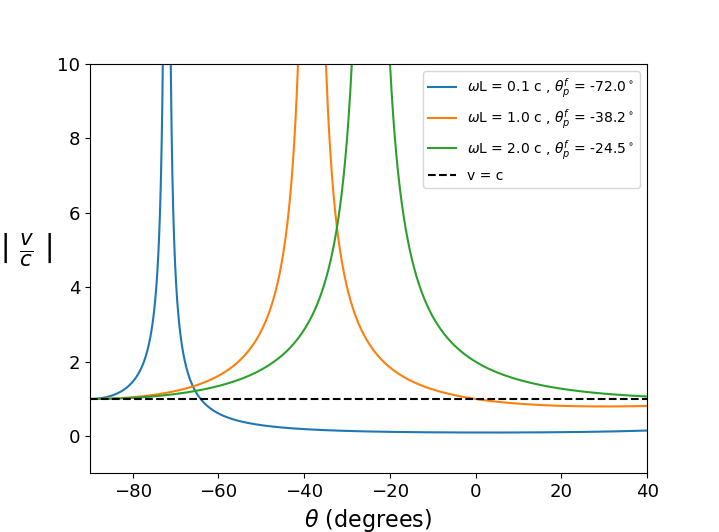}
            \caption{(Case III B) Variation of $\mid \frac{v}{c}\mid$ vs angle $\theta$ of the laser scissors vertex.}
            \label{fig:11}
    \end{figure}

    From the complexity of the graph, it is clear that this case is conceptually distinct. Perhaps the first surprise is the first place the paper is cut. Although the laser points first at a location infinitely far down the $x$ axis, the first point on the paper which is cut is not $\infty$ because it will take an {\it infinite} amount of time for laser photons to reach that far. The angle where the paper is first cut is determined by the  $\theta$ where $d t_{total} / d\theta = 0$. This creates a hole in the paper with a location found from  
    \begin{equation}
        \frac{\sin \theta}{\cos^{2} \theta} = \frac{-c}{L\ \omega} .
    \end{equation}
    The solution is 
    \begin{equation}
        \theta^f_p = \arcsin \Bigg( \frac{L w}{2 c} - \sqrt{\frac{L^{2}\ w^{2}}{4\ c^{2}}+1}\Bigg) .
    \end{equation}
    The plot of $\mid \frac{v}{c}\mid$ vs angle $\theta$ of the laser beam with respect to the vertical is shown in Figure~\ref{fig:11} .
    
    After the first time the paper is cut, there are {\it two} locations \cite{Nemiroff1} on the paper that are associated with each one $t_{total}$ time in Eq. (11). These two locations depict the paper being cut in two directions at once -- both toward the origin at $x=0$, and also toward $x = \infty$. Two different vertices move at once! All of the photons come from the laser, but starting from the initial hole in the paper, those photons that cut the paper toward the origin are emitted after ($t_{rotate}$ is greater) the photons that cut the paper toward infinity. Furthermore, the set of photons cutting the paper towards the origin have a shorter travel time ($t_{traverse}$ is lesser) -- from laser to paper -- than the other set of photons cutting toward infinity. This type of superluminal pair splitting has been verified experimentally in another context \cite{Clericie1501691}.
    
    The vertex cut that moves toward the origin starts with formally infinite speed but then slows. Whether its speed drops below $c$ before reaching the origin depends on $\omega$ and $L$. The vertex cut that moves away from the origin also starts cutting the paper with formally infinite speed, and then it also slows down\cite{Faccio_2018}. The speed of this vertex, though, always approaches $c$ as it moves toward infinite $x$, and is always greater than $c$.

\section{\label{sec:level1}SUMMARY AND DISCUSSION}

It has been shown that the popular physics legend of scissors being able to cut paper faster than light is actually true, in theory, for several simple cases. A scissors vertex moving superluminally does not violate special relativity because it carries no energy or momentum. Therefore, the rip in the paper could start or stop at any time with no change in the energy or momentum to the scissors blades or the paper, and neither of these motional parameters would diverge as the rip approached $c$.  

In the first case, it was shown that when the rotation speed $\omega$ of the upper scissors blade is constant along its infinite length, then for it to close in a finite time, the speed of the vertex between the two scissors blades must approach infinity -- which exceeds $c$. This case is unphysical for rigid scissors, however, if one assumes that the information that the upper blade is rotating moves out from the hinge. This is because this information -- and the force it carries -- can only move along the upper blade at speed $c$ at the most. It will therefore not be possible for the upper blade to remain rigid -- it must either flex or break \cite{web:lang:stats7}.

However, it is possible that the upper blade's motion does not only originate from a single torque applied around the hinge. It could be that the upper blade's angular motion derives from forces distributed along its entire length. If so, then the previous objection will not apply. However, even if the upper blade's rotation arises from a distributed force, it will still not be possible for any part of it to move faster than $c$ relative to the paper and the hinge. For a rigidly rotating upper blade, at some distance $d > c/ \omega$ from the hinge, the blade past $d$ must move faster than light to remain rigid, which is unphysical. However, if $d < x_c$, as defined in Eq. (\ref{xc}), then it is possible for a rigid scissors to cut paper superluminally. 

In the second case of guillotine scissors, when the angle between the blades was small enough, and the downward speed was high enough, the vertex can move -- and the paper can be cut -- superluminally. However, in this case, the guillotine angle at which paper is cut superluminally is actually so small that atomic vibrations and material imperfections in the guillotine blade do not allow this case to be realized practically. 

The third and fourth cases involve cutting the paper with a laser. In the third case, the initial vertex speed scales linearly with the angular speed with which the laser turns, as well as the distance between the swivel point and the paper. Therefore, since these parameters are not limited, the vertex speed can be arbitrarily fast -- even exceeding $c$. Regardless of the initial speed, it was shown that the vertex speed will approach $c$ infinitely far down the paper. 

The last case gives the most interesting results. Here, a hole first forms in the paper at the point of the initial laser contact. From this initial hole, the paper rips outward in two directions simultaneously. Each vertex moves superluminally to start, although the speed of cutting may drop to subluminal. A realistic caveat is that far down the paper the laser photons will become too sparse to cut the paper continuously. 

In sum, it is of educational interest that such a common object as a scissors can, in theory, display such an uncommon attribute as superluminal motion. This counter-intuitive behavior does not violate special relativity and is derivable from straightforward kinematics and Euclidean geometry prevalent in undergraduate physics curricula.

\begin{acknowledgments}
The authors thank Lucas Simonson for initial comments, and Jacek Borysow and an anonymous referee for a critical reading of the manuscript.
\end{acknowledgments}

\section*{\label{sec:level1}REFERENCES}

\nocite{*}
\bibliography{refs}

\end{document}